\documentclass[11pt]{article}
\usepackage{amsmath}
\usepackage{color}
\usepackage{txfonts}
\definecolor{indigo}{RGB}{0,0,120}
\usepackage[colorlinks=true, linkcolor=indigo, citecolor=blue, urlcolor=indigo]{hyperref}

\newcommand{\beq}{\begin{equation}}
\newcommand{\eeq}{\end{equation}}
\newcommand{\beqs}{\begin{eqnarray}}
\newcommand{\eeqs}{\end{eqnarray}}

\usepackage{titlesec}
\titleformat{\section}{\normalsize\bfseries}{\thesection}{1em}{}
\titleformat{\subsection}{\small\bfseries}{\thesubsection}{1em}{}
\titleformat{\subsubsection}{\small\bfseries}{\thesubsubsection}{1em}{}

\begin{document}
%%-----------------------------

\title{A critique of recent semi-classical spin-half quantum plasma theories}

\author{\sc G. S. Krishnaswami$^{1}$, R. Nityananda$^{2}$, A. Sen$^{3}$ and A. Thyagaraja$^{4}$
\\ \\ \small
$^{1}$Chennai Mathematical Institute,  SIPCOT IT Park, Siruseri 603103, India\\ \small
$^{2}$Azim Premji University, PES South Campus, Electronics City, Bangalore 560100, India \\ \small
$^{3}$Institute for Plasma Research, Bhat, Gandhinagar 382428, India\\ \small
$^{4}$Astrophysics Group, University of Bristol, Bristol, BS8 1TL, UK
\\ \\ \small
 Email: {\tt govind@cmi.ac.in, rajaram.nityananda@apu.edu.in, } \\ \small {\tt senabhijit@gmail.com, athyagaraja@gmail.com }}

\date{30 August, 2014}

\maketitle

\begin{abstract}
Certain recent semi-classical theories of spin-half quantum plasmas are examined with regard to their internal consistency, physical applicability and relevance to fusion, astrophysical and condensed matter plasmas. It is shown that the derivations and some of the results obtained in these theories are internally inconsistent and contradict well-established principles of quantum and statistical mechanics, especially in their treatment of fermions and spin. Claims of large semi-classical effects of spin magnetic moments that could dominate the plasma dynamics are found to be invalid both for single-particles and collectively. Larmor moments dominate at high temperature while spin moments cancel due to Pauli blocking at low temperatures. Explicit numerical estimates from a variety of plasmas are provided to demonstrate that spin effects are indeed much smaller than many neglected classical effects. The analysis presented suggests that the aforementioned `Spin Quantum Hydrodynamic' theories are not relevant to conventional laboratory or astrophysical plasmas.

\end{abstract}


{\bf PACS:} 52.35.-g, 52.35.Hr; 52.35.We; 67.10.-j; 67.30.hj

%--------------------------------
\section{Introduction}
%--------------------------------

In recent years attempts have been made to incorporate quantum mechanical concepts like electron spin in plasma dynamics \cite{MarklundBrodin,MahajanAsenjo,BraunAsenjoMahajan,ShuklaEliassonRMP,ShuklaEliassonPRL,Manfredi}. These works seek to extend the governing equations of the classical fluid or kinetic description of plasmas by including quantum effects. They go on to consider the changes to the dynamical description provided by standard classical kinetic theory \cite{LifshitzPitaevski} or fluid models like MHD \cite{Freidberg} or two-fluid models \cite{Hazeltine-Meiss}. Although the resultant literature on this `quantum hydrodynamics' [hereinafter referred to as QHD or SQHD, for those theories involving intrinsic electron spin] is voluminous, there are very few predictions referring to specific, experimentally observable situations where  a clear distinction can be made between them and those of classical plasma models. Furthermore, recently several critiques \cite{VranjesPandeyPoedts,Bonitz,Khan-Bonitz-QHD} have appeared in the literature, pointing to weaknesses in QHD. We did find a striking prediction based on an application of SQHD by Braun et. al. \cite{BraunAsenjoMahajan}. However, we showed in the Comment \cite{critique-arxiv,PRL-comment} that the claims made by these authors of an instability of electromagnetic waves in a cold metallic plasma with a specially prepared spin distribution of electrons, were incompatible with known and experimentally verified properties of electron gases in metals.

The purpose of the present work is to expand on our Comment \cite{PRL-comment} and present a number of examples and arguments which suggest that SQHD in particular has little or no relevance to plasma physics as it is usually applied in laboratory or astrophysical situations. We do not deal with truly quantum plasmas which have to be treated with methods of many-body quantum theory based on powerful tools like Greens functions, Master equations, density functionals and the like. The discussion here is strictly limited to semi-classical descriptions of quasi-neutral plasmas and to enquire whether quantum effects of electron intrinsic spin play a significant dynamical role.

%-------------------------------
\section{Single-particle dynamics}
%-------------------------------

The classical motion of an electron in known electromagnetic and gravitational fields, neglecting radiation reaction, is usually described in two equivalent ways: there is the standard Einstein-Lorentz-Newton formulation incorporating special relativity:
	\beq
	\frac{d{\bf p}}{dt}
      =  -e[{\bf E}+{\bf v \times B}]+{\bf f}
      \quad \text{and} \quad
	\frac{d{\bf x}}{dt} =  {\bf v}.
	\eeq
Here ${\bf p}=m_{e}\gamma {\bf v}$, $\gamma=[1-\frac{v^{2}}{c^{2}}]^{-1/2}$ and $c$ is the speed of light in vacuo. ${\bf E, B}$ are the electric and magnetic fields, assumed to be functions of ${\bf x}$ and the time $t$; ${\bf f}$ is the force of gravity calculated at the position of the electron, which is assumed to have a rest mass $m_{e}$ and charge $-e$.

The same dynamics can be obtained via Hamilton's principle using the Lagrangian:
	\beq
		{\cal L} = m_{e}c^{2}\gamma+e[\Phi-{\bf A.v}]+m_{e}\phi_{g}
	\eeq
where the potentials $\Phi, {\bf A}$ are related as usual to the fields through, ${\bf B} = \nabla \times {\bf A}; {\bf E}=-\nabla \Phi -\frac{\partial {\bf A}}{\partial t}$. Here, $\phi_{g}$ is the gravitational potential, such that, ${\bf f}=m_{e}\nabla \phi_{g}=m_{e}{\bf g}$, where ${\bf g}$ is the acceleration due to gravity on the electron. It has been known from the classical works of Larmor, Alfv\'{e}n and many others that so long as the electromagnetic fields vary slowly relative to the Larmor (cyclotron) frequency $\omega_{ce}=-\frac{eB}{m_{e}}$ and the Larmor radius, $\rho_{e}=\frac{c_{\perp}}{\omega_{ce}}$, the classical motion is described by ``drift orbit theory''. Specifically, if the time rate of change of the fields is measured by the frequency $\omega$ and the spatial scales are represented by the wave number $k$, drift orbit theory may be used when $\rho_{*}= {\rm Max} [k\rho_{e},\frac{\omega}{\omega_{ce}}] \ll 1$. Here, $c_{\perp}$ is the ``peculiar velocity'' of the electron's Larmor gyro motion. In addition, the electron experiences the ``electric drift'' [${\bf V}_{E}=\frac{{\bf E \times B}}{B^{2}}$] and possesses an {\it adiabatic invariant} $\mu_{e}$. In the non-relativistic limit, this is well-known, in leading order [in $\rho_{*}$], to be given by the formula, $\mu_{e}=\frac{1}{2}\frac{m_{e}c_{\perp}^{2}}{B}$. We reproduce these standard formulae to clarify our notations and keep our exposition self-contained. 

It is a consequence of drift orbit theory that $\mu_{e}$ is the ``Larmor'' magnetic moment associated with the electron's gyromotion. The force associated with the Larmor motion is effectively described by the potential $\mu_{e}B$, where $\mu_{e}$ is the adiabatic invariant and $B$ the magnetic field at the position of the electron's guiding centre. The latter moves along the magnetic field with a velocity $v_{\parallel}$ and also drifts perpendicular to the field under the ${\bf V}_{E}$ drift and the ``grad-B'' drift,  given by ${\bf V}_{\nabla B}=\frac{\mu_{e}}{e}\frac{\nabla B \times {\bf B}}{B^{2}}$. If a gravitational field ${\bf f}=m_{e}{\bf g}$ is also present, it would set up a corresponding ``gravitational drift'', ${\bf V}_{g}=-\frac{m_{e}}{e}{\bf g \times B}/B^{2}$ of the guiding centre. The parallel component of the Larmor moment force, $-\mu_{e}\nabla B$ is responsible for the well-known phenomenon of {\it particle trapping} in this Larmor potential, which always arises in an inhomogeneous magnetic field.

Let us consider now the effect of taking into account the intrinsic magnetic moment of the electron, in the spirit of the quantum hydrodynamicists. According to Dirac's relativistic quantum theory, the electron has spin angular momentum projections $\sigma_{\pm}=\pm \hbar/2 \equiv \hbar s_{\pm}$ parallel/anti-parallel to the magnetic field. The theory [with radiative corrections] leads to the intrinsic electron magnetic moment, $m_{\sigma}=-g_{0}\mu_{\rm B} s_{\pm}$, where, $\mu_{\rm B}=\frac{e\hbar}{2m_{e}} = 9.4 \times 10^{-24}$ A.m$^2$ is the Bohr magneton and $g_0$ the gyromagnetic factor ($g_{0} =2[1+\frac{e^2}{4\pi \epsilon_0 \hbar c}+ \ldots]$). As a result, the electron ``feels'' [in a semi-classical picture which SQHD embraces] a force, $-m_{\sigma}\nabla B$. 

We are now in a position to compare the classical [i.e., orbital] Larmor moment and the quantum mechanical intrinsic spin moment:
	\begin{equation}
	\frac{|m_{\sigma}|}{\mu_{e}} 
      \simeq \frac{\mu_{\rm B}}{\mu_{e}}
      \simeq \frac{\mu_{\rm B}B}{\frac{1}{2}m_{e}c_{\perp}^{2}} 
      \simeq  \frac{\hbar \omega_{ce}}{T},
	\end{equation}
where we assume that the electron is a ``typical'' one in a plasma at a local temperature, $T$ [in joules; $1$keV $\simeq 1.6\times 10^{-16}$ joules].  

The above considerations lead to the first simple estimate of the relative sizes of the forces on the electron due to the Larmor and intrinsic moments. As an example, consider a typical fusion plasma in magnetic confinement  (as in the JET tokamak): here, typically $T \simeq 10$keV, $B \simeq 10$T, $\hbar =10^{-34}$Js, $m_e \simeq 9 \times 10^{-31}$kg, $e \simeq 1.6 \times 10^{-19}$C. Substitution gives, $\omega_{ce} \simeq 1.8 \times 10^{12}$rads/s; $\hbar \omega_{ce} \simeq 1.8 \times 10^{-22}$J. Hence, $m_{\sigma}/\mu_{e} \simeq 10^{-7}$ and the spin magnetic moment of the electron is seven orders of magnitude smaller than its Larmor gyro moment. Thus the SQHD effects are totally negligible in the individual electron equation of motion, even when that electron is maximally [namely, exactly] polarised along [ie., parallel or anti-parallel to] the equilibrium field. Furthermore, whereas the spin orientations of the various electrons can cancel, the Larmor motions of {\it all} electrons, irrespective of their thermal energies, are always in the diamagnetic direction. It is therefore abundantly clear that in treating classical, magnetically confined fusion plasmas, one can totally neglect quantum/intrinsic spin effects. This estimate implies that many other important higher order effects of classical plasmas [effects of collisions, relativistic corrections, polarization drifts, radiative corrections] will generally totally overwhelm the so-called ``spin quantum effect'', and must therefore be considered along with it in any consistent physical model.

As a second example, consider an inter-galactic cluster plasma in the vicinity of an AGN. Then, the ambient magnetic field $B_{0} \simeq 10^{-9}$T, while the electron number density and temperature are $n_{e} \simeq 10^{6}$m$^{-3}$, $T_{e} \simeq 1-10$keV. Very large electron accelerations are thought to occur due to Alfv\'{e}n waves in the tenuous, relatively hot, non-relativistic plasma generated by strong flows associated with the AGN. Here $\omega_{pe} \simeq 5\times 10^{4}$rad/s, $\omega_{ce} \simeq 176$rad/s. Thus, $\frac{\hbar \omega_{ce}}{T} \leq 10^{-16}$ and spin quantum effects are highly suppressed.

The above estimates are essentially similar to one arrived at by Marklund and Brodin \cite{MarklundBrodin}. They went on to suggest that perhaps the spin terms could be more important when the Larmor moments are smaller. It is plain that this requires the electron to be ``cold'' in the sense $T \simeq \hbar \omega_{ce}$. At $B \simeq 10$T, the temperature has to be $T \simeq 1.8 \times 10^{-22}$J, i.e. $\simeq 13$K. The plasma would have to be far cooler still at lower fields. It is therefore easy to demonstrate that SQHD  cannot be applied to classical, Maxwellian plasmas at temperatures and densities typical of such systems. We note that Vranjes et. al. \cite{VranjesPandeyPoedts} have directed similar criticism at QHD. 

It is instructive to take a brief look at the quantum mechanical formulation of electrons [with spin] moving in a magnetic field. We note that the well-known non-relativistic formula [in SI units] for the energy levels (`Landau levels') of an electron in a magnetic field is,
	\beq
	E_n = \left(n+\frac{1}{2}+\sigma \right) \hbar \frac{|e|B}{m_{e}}+\frac{p_z^2}{2m_e}
	\eeq
[Eq.(125.7) of \cite{LandauLifshitzQM}], where $\sigma= \pm \frac{1}{2}$. When $T \gg \hbar \omega_{ce}$, the quantum number $n$ would be very large, and the tiny spin-dependent correction to the ``orbital energy'' contributed by the high $n$ term is clearly an insignificant effect. Any semi-classical approach must require $n \gg 1$ and thus cannot possibly account for the spin quantum number $\sigma$ or effects arising from it. Thus, the quantum [Pauli equation] and the classical estimates based on the Lorentz-Einstein equations agree that intrinsic spin effects are truly sub-dominant to many other effects which must be accounted for. For example, for $T \geq 1$keV $\approx 10^7$K, the relativistic corrections to the electron mass imply kinetic energy changes much larger than the spin-quantum energy. This can be seen from the fact that with $B=1$ Tesla, $\frac{2T}{m_{e}c^{2}} \gg \frac{\mu_{\rm B}B}{T}$. Chandrasekhar's white dwarf theory [where electron spin energies are neglected] clearly depends upon the presence of relativistic electrons at the Fermi level to obtain the relativistic equation of state.

%---------------------------------
\section{Critique of fluid formulations of spin quantum plasmas}
%---------------------------------

In this section, we point out specific problems with the spin quantum plasma formalism developed by Marklund and Brodin and others (see \cite{MarklundBrodin} and the review \cite{ShuklaEliassonRMP}) and applied by Mahajan et. al. \cite{MahajanAsenjo,BraunAsenjoMahajan} to propose a `spin-laser'. 

%---------------------------------
\subsection{ Fermi-Dirac statistics and the spin quantum plasma formalism}
%---------------------------------

We consider first the paper by Marklund and Brodin\cite{MarklundBrodin}, as it appears to form the basis for subsequent works in this area. The authors make the explicit claim: ``In this Letter, we present for the first time the fully nonlinear governing equations for spin-1/2 quantum electron plasmas. Starting from the Pauli equation describing the non relativistic electrons, we show that the electron-ion plasma equations are subject to spin-related terms. These terms give rise to a multitude of collective effects, of which,  some are investigated in detail. Applications of the governing equations are discussed, and it is shown that under certain circumstances {\it the collective spin effects can dominate the plasma dynamics.}''[our italics]. 

We have two major (related) concerns with the theory of spin quantum plasmas, as presented in the papers cited. Firstly, the derivation of the ``spin quantum plasma equations'' in \cite{MarklundBrodin} starts with the independent electron approximation in which the Coulomb interaction of the electron gas is neglected and the single-electron, non-relativistic Pauli equation is invoked [cf. Eq.(1) of \cite{MarklundBrodin}]. This assumption was also made by Sommerfeld in his theory of the free-electron gas in a metal \cite{LifshitzPitaevski,AshcroftMermin,Kittel,RASmith,Ziman,KHuang}. However, in the paragraph preceding this equation, the authors state that they assume [unlike Sommerfeld who explicitly invokes Fermi-Dirac(FD) statistics for the electron gas] the simple product representation of the $N$-electron wave function of the system [rather than the correct, Slater-determinantal wave functions incorporating FD statistics]. They state, apparently in justification: ``Thus, we will here neglect the effects of entanglement and focus on the collective properties of the quantum electron plasma''. In our view, it is a serious error to ignore anti-symmetrization of the many-electron wave function and FD statistics. This plainly contradicts Pauli's exclusion principle.

The authors' neglect of the special type of electronic correlation  implied by the exclusion principle, especially when dealing with any many-identical-fermion system when the temperature is well below the Fermi temperature [defined for example in \cite{AshcroftMermin,Kittel,RASmith}]
	\beq
	T_F = \frac{\hbar^2}{2 m_e} \left(3 \pi^2 n_e \right)^{2/3}
	\eeq
where $n_e$ is the electron number density, renders their theory inconsistent with known experimental facts in this regime. This neglect is contrary to the fundamental principle of quantum statistics, that when the electron thermal de Broglie wavelength 
	\beq
	\lambda=\frac{\hbar\sqrt{2\pi}}{(m_{e}T)^{1/2}}
	\eeq
is of order or larger than the inter electron distance [$n_{e}^{-1/3}$, see \cite{KHuang}, p. 226], the exclusion principle constraints and FD statistics are essential. Thus, we must distinguish between $T \leq T_F$ `quantum plasmas', which must be described according to quantum many-body theory [and shown to be fully consistent with FD statistics, Fermi liquid theory, quantum Master equations etc.] and `fully ionized classical plasmas' with $T \gg T_F$, and which are described in local thermodynamic equilibrium by Maxwell-Boltzmann distributions. Marklund and Brodin's use of the simple product wave function is only acceptable for {\it distinguishable fermions} or when the Fermi gas is so hot that Maxwell-Boltzmann(MB) statistics applies - as it does in fusion plasmas, for example, where MB distributions describe thermodynamic equilibria and approximate local thermodynamic equilibrium under collisional conditions [cf.\cite{RASmith},p.42]. This can happen only when $\lambda \ll n_e^{-1/3}$. When this condition fails to hold, it is essential to use Slater determinantal wave functions or a second-quantized formalism in working out all {\it average, fluid properties} of the electron gas \cite{LifshitzPitaevski,AshcroftMermin}. The failure to do so can result in some strange properties being assigned to the electron gas, at variance with both standard theory and experiments. It is well-known\cite{LifshitzPitaevski,AshcroftMermin,Kittel,RASmith,Ziman} that the failure of Drude's classical electron theory of metals and negligible electronic contribution to the specific heat and magnetic properties (such as the smallness of Pauli spin-paramagnetism and Landau diamagnetism) are direct consequences of FD statistics of the electron gas.  As noted in \cite{Pake}, the effects of electron spin (specifically electron paramagnetic resonance) are strongly suppressed in the conduction electron gas by the very small paramagnetic (Pauli+Landau) susceptibility.

Secondly, we note that an immediate consequence following from the authors' neglect  of FD statistics for the electron gas is that, their formula for the magnetization of the electron plasma- treating, as they do, the ion fluid as a uniform neutralizing background for simplicity- is grossly in error. According to them, the magnetization spin current is,
	\beq
	\label{jspin}
	{\bf j}_{\rm sp} = \nabla \times [2 n_e \mu_{B}{\bf S}]
	\eeq
Here, $n_e$ is the conduction electron number density and ${\bf S}$ is the local average ``spin vector'' of the electrons at the elementary volume over which $n_e$ is reasonably constant. Now, we should stress that we are only interested in magnetized plasmas, since the spin effects are relevant [if at all] only in these systems. Marklund and Brodin identify $\mu_{B} B_{0}/T$ as a dimensionless measure of quantum effects. Here $B_0$ is the ambient external magnetic field and $T$, the electron temperature in Joules. The numerator is evidently the intrinsic electron spin-magnetic moment energy in the external field. As mentioned in the previous section, the classical thermal Larmor gyro motion of the electron in a magnetic field is associated with an adiabatic invariant Larmor gyromagnetic moment $\mu_e$ such that $\mu_{e}B_{0}=E_{\perp}=\frac{2}{3}E \simeq T$. Hence, the parameter in question is simply, $\mu_{B}/\mu_{e}$. Marklund and Brodin correctly state that in high temperature plasmas or in the presence of significant fields, this parameter is small and hence ``spin quantum effects'' are negligible in comparison with the usual gyromagnetic moment effects. What they {\it fail} to mention is the existence and physical significance of the degeneracy parameter, $D=T/T_{\rm F}$. In order to apply MB statistics, it is essential that $D \gg 1$\cite{RASmith}. This degeneracy condition for FD statistics is equivalent to the principle stated earlier; thus, $D\gg 1$ implies, $\lambda \ll n^{-1/3}$.  In metals, with $n \simeq 10^{28}$ to $10^{29}$m$^{-3}$, $T_{\rm F}$ is a few  eV [cf. Table 1.1 and Table 2.1 in \cite{AshcroftMermin}]. Hence at room temperature, the conduction electron gas is degenerate [$D \simeq 10^{-2}$] and the authors' claim that at ``low temperatures'' quantum spin effects could be important is essentially incorrect. By Pauli ``blocking'', the spin magnetization should be of order $2(n_{+}-n_{-})\mu_{B}{\bf b}$, where ${\bf b}$ is the unit vector in the local magnetic field direction and $n_{\pm}=\frac{n}{2}[1 \pm \frac{3\mu_{B}B_{0}}{2T_{\rm F}}]$. Here $n_{\pm}$ refer respectively to the number of electrons per unit volume with their intrinsic spin vectors parallel and antiparallel to the local magnetic field. Note that $n_+ + n_- =n$ and $\frac{n_{+}-n_{-}}{n}=O[\frac{T}{T_{\rm F}}]$  [cf. \cite{RASmith,Ziman,KHuang}]. Physically this means that at temperatures low compared to the Fermi temperature, which depends solely on the electron number density and physical constants, only [$\sim n(\frac{T}{T_{\rm F}})$] electrons in a layer close to the Fermi level contribute to ${\bf j}_{\rm sp}$ \cite{AshcroftMermin,Kittel}. This is also a direct consequence of the kinetic theory of Fermi liquids [cf. \cite{LifshitzPitaevski}] from which any reasonable {\it fluid} theory of the electron plasma ought to be derived using an appropriate Chapman-Enskog asymptotic expansion in powers of the relevant Knudsen number (inverse collisionality parameter measuring the departure from local thermodynamic equilibrium \cite{LifshitzPitaevski,Ziman}).

Thus, their formula for ${\bf j}_{\rm sp}$ greatly over-estimates the spin magnetization current when $T \ll T_F$ [see below for numerical estimates]. On the other hand, when $T\gg T_{F}$ and MB statistics apply [as happens in all fusion and astrophysical plasmas except in white dwarf cores], the quantum intrinsic spin effects are entirely negligible both on individual electrons and collectively, as argued in the previous section.

%------------------------------------
\subsection{Inconsistent use of Pauli blocking in spin-laser prediction}
%------------------------------------

We now come to the paper by Braun et. al. \cite{BraunAsenjoMahajan} which proposes light amplification driven by inhomogeneous quantum spin fields in low-temperature conduction electron plasmas in metals. The authors argue that an EM wave entering a metal at low temperatures ($\sim 30$K) with a suitably prepared internal spin field $\bf S$ with large gradients, would suffer an instability and be amplified. This work was based on the quantum spin vorticity formalism of Asenjo and Mahajan \cite{MahajanAsenjo}, which is in turn based on the quantum spin plasma equations of Marklund and Brodin \cite{MarklundBrodin}.

However, these papers suffer from contradictory assumptions in dealing with the electron gas in a metal, as pointed out in our Comment \cite{PRL-comment}. On the one hand, the authors explicitly assume that the ``average intrinsic spin vector'' ${\bf S}$ is a unit vector. Our analysis shows that $\bf S$ must be of order $[\frac{n_{+}-n_{-}}{n}]{\bf b} \ll 1$, where ${\bf b}$ is a unit vector. The assumption that ${\bf S}$ is a unit vector leads to a very large estimate for the ``spin magnetisation current density'' given by the Marklund-Brodin formula Eq.(\ref{jspin}) quoted above. Thus, we may estimate, $j_{\rm sp} \simeq  2n\frac{\mu_{B}}{L_{S}}\simeq 10^{5}$ to $10^{6}{\rm A.m}^{-2}$, where, $L_{S}$ is the gradient length-scale of the spin field.  We note that equilibrium conditions require that this should also be of the same order as the density length-scale. The above estimate corresponds to a deliberately chosen ``macroscopic'' scale of 1m. Any smaller choice [e.g. $L_{S} \simeq 10^{-2}$m] will make the current density estimate even larger! This $j_{\rm sp}$ corresponds to a magnetic induction of $0.1-1$ Tesla. Such a large magnetic induction is more common for electrons in a ferromagnet than for conduction electrons in a metal. A saturation magnetization for the degenerate conduction electron gas would correspond to creating a highly excited state. At $T \simeq 0$, the energy per unit volume required to create a totally ``spin polarized'' electron gas by flipping half the spins and promoting them to states above the Fermi level is roughly of order $(n/2) T_F$. Thus, if we take $n \simeq 10^{28} -10^{34}$m$^{-3}$, and $T_F \geq 1$ eV, the energy density of the plasma would be of order, $10^{9}-10^{15}$ Joules/m$^3$! The magnetic field required for such a large polarization must then be $B \simeq T_F/\mu_{\rm B} \simeq 10^{4}$T!

On the other hand, Braun et. al. assume a very low temperature $T \simeq 30$K, in order to prevent electron-electron and electron-phonon collisions from totally damping the electromagnetic wave that moves into their medium. At low temperatures, the low collision frequency is due essentially to Pauli blocking, which was not taken into account in their equations. The problem is that their formalism and equations are valid neither at high temperatures nor at low temperatures. At high temperatures Larmor gyromagnetic moments will dominate over quantum spin effects and Coulomb collisions imply very short mean-free paths at high densities. At low temperatures, the exclusion principle and Pauli blocking greatly reduce the spin magnetization current and make the assumption of a unit ${\bf S}$ invalid.

In their reply \cite{PRL-comment-reply} to our Comment, the authors of \cite{BraunAsenjoMahajan} accept the validity of our criticisms. In particular, they agree that (1) the average spin field $\bf S$ is not a unit vector and (2) that Pauli blocking (PB) will greatly reduce the spin magnetization current in low temperature metallic plasmas, and thereby reduce any instability. They go on to state that at low temperatures, the light wave growth rate $\Gamma$, as well as electron-electron collision/damping frequency $\nu_{ee}$, will both be brought down by the factor $\alpha = T/T_F$, yielding
	\begin{equation}
	\Gamma_{\rm PB} = \alpha \Gamma_{old} \quad \text{and} \quad
	\nu_{ee} \sim \frac{k_B T^2}{\hbar T_F}.
	\end{equation}
Thus they estimate the ratio of collisional damping to growth rate as
	\begin{equation}
	\frac{\nu_{ee}}{\Gamma_{\rm PB}} \sim \frac{k_B T}{\hbar \Gamma_{old}}.
	\end{equation}
They assert that, in principle, this ratio could be less than unity for sufficiently low temperatures, thereby implying amplification of the light wave.

It is true that in an `ideal metal', the electron collision rate $\nu_{ee}$ will scale like $T^{2}$ [cf. \cite{LifshitzPitaevski}, also Eq.(17.66) in \cite{AshcroftMermin}]. In reality, at very low temperatures [when, $T \ll T_{\rm Debye}$], the `residual resistance' of a metal due to impurity or lattice defect scattering leads to a temperature-independent collision rate $\nu_0$. If such a temperature-independent collision frequency is used, then the authors' estimate for the ratio of collision rate to growth rate $\frac{\nu_0}{\Gamma_{\rm PB}} \sim \frac{T_F \nu_0}{T \Gamma_{\rm old}}$ would be  {\it more} than unity for sufficiently low temperatures, implying that collisions prevent any light amplification.

Even if we accept $\nu_{ee} = k_B T^2/\hbar T_F$ as a reasonable low temperature collision rate, as well as their ad hoc method of estimating the effect of Pauli blocking [i.e., to simply assume that $\Gamma_{\rm PB} = \frac{T}{T_F} \Gamma_{\rm old}$], we find for solid state plasmas, that the ratio of collision to growth rate $\nu_{ee}/\Gamma_{\rm PB}$ is less than unity only for very low temperatures $T < 0.025 K$. Collisional damping of the wave will overwhelm the claimed effect for any higher temperature. What is more, even at such a low temperature, the EM wave would have to travel at least $c/\Gamma_{\rm PB} \sim 30$ km in their medium to be significantly amplified. To obtain these estimates for solid state plasmas considered by the authors, we take $n_e \approx 10^{29}$m$^{-3}$, corresponding to a Fermi temperature $k_B T_F \approx 1$ eV $\approx 10^4$ K and a plasma frequency $\omega_{pe} \approx 1.6 \times 10^{16}$ s$^{-1}$. Now from Fig. 1 of their Letter \cite{BraunAsenjoMahajan}, the {\it maximum} value of $\Gamma_{\rm old}$ is $2.5 \times 10^{-7} \times \omega_{pe} \approx 4 \times 10^{9}$ s$^{-1}$. Using their formulas $\nu_{ee} \sim \frac{k_B T^2}{\hbar T_F}$ and $\frac{\nu_{ee}}{\Gamma_{\rm PB}} \sim \frac{k_B T}{\hbar \Gamma_{\rm old}}$, we find that $k_B T < 2.5 \times 10^{-6}$ eV $\approx 0.025$K for the growth rate to exceed the collision rate. At $T = 0.025$ K, $\Gamma_{\rm PB} \approx 10^4$ s$^{-1}$ and so the wave must travel $c/\Gamma_{\rm PB} \approx 30$ km in the medium to be amplified significantly. Thus, even assuming their formalism to be corrected as suggested by the authors, the effects predicted are negligible [using their own numbers and formulae] and are far smaller than many other neglected effects such as collisionless damping, impurity scattering, etc.

%------------------------------
\subsection{Irrelevance of Bohm force in the fluid approximation}
%------------------------------

In addition to forces arising from the intrinsic spin of electrons, the quantum hydrodynamicists also introduce a quantum Bohm force ${\bf F}_Q$ in their electron fluid momentum equation [see Eqs.(17,27) of the review by Shukla and Eliasson \cite{ShuklaEliassonRMP}]
	\beq
	m_e \left( \frac{\partial \bf u}{\partial t} + {\bf u}\cdot \nabla {\bf u} \right) = - \frac{1}{n_e} \nabla P_F + e \nabla \phi + {\bf F}_Q,
	\quad \text{where} \quad
	{\bf F}_Q = \frac{\hbar^2}{2 m_e} \nabla \left( \frac{\nabla^2 \sqrt{n_e}}{\sqrt{n_e}} \right).
	\eeq
However, fluid-like models can only be applied when the wave length of any perturbation is significantly large compared with the inter-particle distance $r_{\rm int}$. A comparison between the pressure gradient terms and the Bohm term reveals that when this condition is satisfied, the Bohm force is negligible and certainly less important than neglected off-diagonal stresses due to interactions. This is demonstrated by the following estimates. Using the ideal gas pressure law $P_F = n_e T_F$,
	\beq
	\left\vert -\frac{1}{n_e}\nabla P_F \right\vert \; \simeq \; 
	k \; \delta T_F \; \simeq \; 
	\frac{\pi^{2/3}}{3m_{e}} \hbar^2 n_e^{-1/3} k \; \delta n_{e} 
	\eeq
where $k$ is a typical wave number and $\delta n_{e}$ is the density perturbation. On the other hand, the Bohm term $\simeq \frac{\hbar^2}{2m_e}k^3 \left(\frac{\delta n_e}{n_e} \right)$. We see that in order for the Bohm term to be comparable with the degenerate Fermi pressure gradient, we must have $1/k \simeq n^{-1/3} \simeq r_{\rm int}$. But when the wavelength of the perturbation is comparable to the inter-particle separation, semi-classical fluid models (or for that matter, kinetic models) cease to apply!

%---------------------------------
\subsection{An ICF example: comparing spin force with electric and Lorentz forces}
%---------------------------------

We have argued that for $T > T_F$ where Maxwell-Boltzmann statistics apply, the effects of quantum mechanical intrinsic spin magnetic moments are unimportant compared to classical/orbital Larmor magnetic moments. Let us examine this, along with the validity of the fluid approximation, in an example of an electron plasma under conditions possibly representative of inertial confinement fusion. This complements our earlier example of hot magnetic fusion plasmas. Suppose $T \simeq 500$eV $\simeq 6 \times 10^7$K, $T_{F} \simeq 100$eV $\simeq 1.2 \times 10^7$ K, $n_{e}\simeq 5\times 10^{30}$m$^{-3}$, $B \simeq 5\times 10^{4}$T. From these data the following parameters may be estimated: the plasma frequency $\omega_{pe} = \sqrt{n_e e^2/\epsilon_0 m_e} \simeq 1.3 \times 10^{17}$rad/s; cyclotron frequency $\omega_{ce} = eB/m_e \simeq  8.8 \times 10^{15}$ rad/s; electron thermal speed $v_{\rm the} = \sqrt{2T/m_e} \simeq  1.3 \times 10^{7}$ m.s$^{-1}$; inter-electron distance $r_{\rm int} = n_e^{-1/3} \simeq 5.8 \times 10^{-11}$m; Debye screening length $\lambda_{\rm Debye} =\frac{v_{\rm the}}{\omega_{p}} \simeq 1.1 \times 10^{-10}$ m; Larmor radius $r_{L} =\frac{v_{\rm the}}{\omega_{ce}} \simeq  1.5 \times 10^{-9}$m; de Broglie wavelength $\lambda_{\rm deBr} = \hbar/m_{e}v_{\rm the} \simeq 8.7 \times 10^{-12}$ m; $n_{e}\lambda_{\rm Debye}^{3}=  5.8$; $\frac{\hbar \omega_{ce}}{T} \simeq 10^{-2}$

These numerical estimates imply that there are merely 5 to 6 electrons in a `Debye cube' so that this may not quite be regarded as a ``collective plasma''. Indeed, the smoothed out self-consistent field approximation is likely to breakdown in this case. However, it may be reasonable to use classical approaches, since $r_{\rm int}>\lambda_{\rm deBr}$, though only marginally so. It is not clear what the electron ``relaxation frequency'', $\nu_{\rm coll}$ is. If we adopt the conservative estimate applicable to metals at lower density of $\nu_{\rm coll} \simeq 10^{14}$s$^{-1}$, the condition that $\omega_{ce} \gg\nu_{\rm coll}$ [required for deriving magnetized plasma fluid equations] is again only marginal. Clearly, at the stated temperature, there is no trace of the lattice structure in the ion plasma. This may allow the usual Landau collision integral to be applied, leading to the Spitzer-Braginskii electron collision rate, (we use $\log \Lambda \approx 10$ for the Coulomb logarithm below)
	\begin{eqnarray}
	\label{eq.9}
	\nu_{\rm Br} 
      & = & \frac{e^4 \: n_e \: \log \Lambda}{6 \sqrt{3} \pi \:
      \epsilon_0^2 \: \sqrt{m} \: T^{3/2}} \approx 10^{16} \: s^{-1}.
	\end{eqnarray} 
This also suggests that $\nu_{\rm coll} \simeq 10^{14}-10^{16}$s$^{-1}$. Even adopting the lower value, we see that the mean-free path due to collisions, $v_{\rm the}/\nu_{\rm coll} \simeq 10^{-7}$m; indeed, it could be as short as the Larmor radius $r_{L}$, if the Spitzer-Braginskii value is used! This implies a plasma resistivity $\eta = \frac{m_{e}\nu_{\rm coll}}{n_{e}e^{2}} \geq 10^{-9}$ohm-m, and a resistive diffusivity, $D_{\rm res} = \frac{\eta}{\mu_{0}} \geq10^{-3}-10^{-1}$m$^{2}$s$^{-1}$. Adopting the Spitzer-Braginskii collision frequency implies that the magnetization of the plasma is not at all complete, since $\omega_{ce} \simeq \nu_{\rm Br}$. We recall that we are interested in magnetized plasmas, since the spin effects are relevant [if at all] only in these systems. It is also clear that any wave which can penetrate this plasma must have frequencies significantly higher than $\omega_{pe}$, the cut-off frequency, i.e., with wave number $k \gg \frac{\omega_{\rm pe}}{c} \simeq 4.3 \times 10^8$ m$^{-1}$.

Now, since the electron motion can be treated classically in these conditions, each electron feels the electromagnetic self-consistent fields and the intrinsic spin magnetic force $\mu_{\rm B}\nabla B$, as well as Coulomb collisions with the ion background [assumed small for the present estimates]. If we set, $|\frac{\delta {\bf B}}{B_{0}}|=\tilde{b}$, the linearized Lorentz force is estimated by $F_{\rm L} \simeq e \, v_{\rm the} \, B_0 \:\tilde{b}$. The spin force is estimated as $F_{\rm sp} \simeq (e\hbar /m_{e})B_{0}k \tilde{b}$, where $k$ is the perturbation wave number. This is an {\it over-estimate} since at the high temperature, the effective magnetic moment for an individual electron must be reduced by the factor, $\hbar \omega_{ce}/T \simeq 10^{-2}$. Then,
	\beq
	\frac{F_{\rm spin}}{F_{\rm L}}
     = \frac{(e\hbar /m_{e})B_{0}k \tilde{b}}{ev_{\rm the}B_{0}\tilde{b}} 
     = \frac{\hbar}{m_{e}v_{\rm the}}k
     = \left(\frac{\lambda_{\rm deBr}}{r_{L}} \right) (kr_{L}).
	\eeq
Plainly, the ``spin term'' is smaller than the Lorentz term by the quantity, $\lambda_{\rm deBr}/r_{L} \approx 6 \times 10^{-3}  \ll 1$, whenever, $kr_{L} \leq 1, k \leq 10^{9}$m$^{-1}$. If waves with this wave number are considered, the dynamics must be treated kinetically as in standard ``hot plasma'' physics where fluid theories are not valid when the wavelengths approach the electron collisionless skin depth $c/\omega_{\rm pe} \simeq 2 \times 10^{-9}$m or Larmor radius $r_{L}$. Furthermore, such short wavelength perturbations are subject to heavy phase-mixing damping mechanisms.

We may compare the spin term with the electric field term, $F_{\rm E} \simeq |e\tilde{E}|$ in the equation of motion. From Faraday's law, we see that $\tilde{E}\simeq (\omega/k)B_{0} \tilde{b}$. This immediately implies that for $\omega \simeq \omega_{ce}; k\simeq 1/r_{L}$, the ratio of the spin term to the electric field term is, $F_{\rm sp}/F_{\rm E} = \frac{\hbar \omega_{ce}}{T} \ll 1$, which is precisely the factor we obtained comparing the Larmor magnetic moment and the intrinsic spin moment! The plain fact remains under the given conditions, the electric field term, the magnetic Lorentz term and the electron's inertia are all comparable in the single-electron Newton-Lorentz equation. The intrinsic spin is a quantum effect and as such is strongly sub-dominant in that equation, since the spin-energy $\mu_{\rm B}B$ is always small compared to the kinetic energy $\simeq T$ in the conditions considered. What is more, the above value of resistive diffusivity $D_{\rm res} \geq 10^{-3} - 10^{-1}$ m$^2/$s (typical of even tokamak plasmas of much lower particle density) indicates that collisions are more important than spin quantum effects in the electron momentum balance equation [generalized Ohm's law].

%-------------------------------------------
\section{Discussion and conclusions}
%-------------------------------------------

The construction of a semi-classical kinetic theory in the spirit of the standard plasma kinetic theories \cite{LifshitzPitaevski} including the intrinsic electron spin $\mu_{\rm B}$ is not straight forward. If we proceed from a Hamiltonian approach in the collisionless case, the electron fluid would have to be described by two distribution functions in ${\bf x,p}$ phase space: thus, $f_{+}({\bf x,p},t)$ describes electrons with spins oriented parallel to the local ${\bf B}$ field and $f_{-}({\bf x,p},t)$ describes the electrons with spins oriented anti-parallel to the field at any instant and location. While the collisionless kinetic equations for these functions are readily written down in terms of the Hamiltonians, 
	\beq
H_{\pm}[{\bf x, p},\Phi, \phi_{g}, {\bf A}]=m_{e}c^{2}\gamma-e[\Phi-{\bf A.v}]-m_{e}\phi_{g} \pm \mu_{B}B,
	\eeq
transitions between these two species must be allowed for. As our estimates in Section I show clearly, where this sort of semi-classical model is physically relevant, the ``quantum spin potentials'' are tiny [like $\frac{\hbar \omega_{ce}}{T}$] compared to the kinetic energy terms. As such, any effects from them should be calculable from a straight-forward perturbation expansion. The corresponding fluid models could, in principle be derived from such semi-classical Fokker-Planck equations [with collisions] using the usual methods expounded in \cite{LifshitzPitaevski}. For colder plasmas typical of condensed matter, only a strictly quantum approach based on Fermi liquid or similar theories based on quantum Boltzmann equations would seem to be the correct approach. For the reasons we have already discussed with various estimates, it is not clear to us what role the intrinsic spin effects are supposed to play. It is possible that SQHD could be re-formulated along these lines and new predictions from it tested against experiments in the appropriate regimes. 

We are not aware of any serious comparison between SQHD [as we understand it from the references cited herein] and extant results in metallic [ie quantum plasmas] such as optical propagation properties, de Haas-van Alphen effect, cyclotron resonance, magneto resistance, quantum Hall effect etc. These well-known and experimentally well-tested effects suggest that semi-classical treatment of the conduction electron permittivity [see eg. \cite{LifshitzPitaevski}, Ch. III] are likely to be grossly in error. Thus SQHD seems not to make any contact with works like those of \cite{Ashcroft1} and Fermi Liquid Theory [cf. \cite{LifshitzPitaevski}] [in the cases when $T\ll T_{F}$]. We would welcome published experimental verifications for SQHD [as formulated in the extant literature] in metallic plasmas.

It is clear that the classical theory of the electron plasma requires that $m_{e}v_{\rm the}/(\hbar k) \gg 1$ and the plasma must be fully ionised. We have not been able to find any systematic appraisal of the regimes when SQHD can be expected to {\it fail} and when it produces new effects in classical plasmas at variance with standard kinetic theory of plasma waves, as expounded for example in \cite{Brambilla} or in the classic work of Stix \cite{Stix}. 

We do not understand the physical bases of the extended Vlasov equations derived in semi-classical SQHD \cite{ZamanianMarklundBrodin}. The fact that the intrinsic electron spin is a discrete and non-continuous degree of freedom with no direct semi-classical limit appears to be violated. Any such Vlasov equation should be analysed carefully to check whether entropy is conserved [as in standard Vlasov kinetics]. It is also important to formulate the Poynting theorem which accounts for the relevant wave-particle interactions and the resultant stable/unstable behaviour of the wave systems [see, for example, \cite{Brambilla} for an excellent discussion of this important point].

In conclusion, in the absence of deeper theoretical or suitable experimental verifications of semi-classical SQHD, this model appears to have no relevant applications to laboratory or astrophysical plasmas of interest. Possibly more work on the models discussed might yield results of real significance and applications of interest. Our discussion suggests that at present there is no strong evidence of this.\\

{\noindent \bf Acknowledgements:} We thank a referee for stimulating and constructive suggestions. The work of GSK was supported by a DST Ramanujan fellowship of the Govt. of India.

%---------------------------------

\end{document}